\documentclass[runningheads,fleqn]{svmult}
\usepackage{makeidx}   % allows index generation
\usepackage{graphicx}  % standard LaTeX graphics tool
\usepackage{subeqnar}  % subnumbers individual equations
\usepackage{multicol}  % used for the two-column index
\usepackage{taphys}    % flushleft layout of captions etc.
\makeindex             % used for the subject index
\begin{document}
\title*{
Quantum Critical Metals: beyond the Order Parameter Fluctuations}
\toctitle{Quantum Critical Metals: beyond the Order Parameter 
Fluctuation Theory}
% allows explicit linebreak for the table of content
%
%
\titlerunning{
Quantum Critical Metals}
% allows abbreviation of title, if the full title is too long
% to fit in the running head
%
\author{Qimiao Si
%%\inst{1}
%%\and Roger Temam\inst{2}
%%\and Jeffrey Dean\inst{2}
%%\and David Grove\inst{1}
%%\and Craig Chambers\inst{2}
%%\and Kim~B.~Bruce\inst{2}
%%\and Elsa Bertino\inst{1}
}
%
%%\authorrunning{Qimiao Si et al.}
\authorrunning{Qimiao Si}
% if there are more than two authors,
% please abbreviate author list for running head
%
%
\institute{
%%Princeton University, Princeton NJ 08544, USA
%%\and Universit\'{e} de Paris-Sud,
%%     Laboratoire d'Analyse Num\'{e}rique,
%%     B\^{a}timent 425,\\
%%     F-91405 Orsay Cedex, France
Department of Physics \& Astronomy, Rice University, Houston,
TX 77005, U.S.A
}

\maketitle              % typesets the title of the contribution

\begin{abstract}
The standard description of quantum critical points takes into account
only fluctuations of the order parameter, and treats quantum fluctuations
as extra dimensions of classical fluctuations. This picture can break down
in a qualitative fashion in quantum critical metals: 
non-Fermi liquid electronic excitations are formed precisely at the 
quantum critical point and appear as a part of the quantum-critical spectrum.
In the case of heavy fermion metals, it has been proposed that
the non-Fermi liquid behavior is characterized by the destruction
of the Kondo effect.
The latter invalidates Hertz's Gaussian theory of paramagnons and 
leads to an interacting theory that is ``locally quantum critical''.
We summarize the theoretical and experimental developments
on the subject. 
We also discuss their broader implications, and make contact with
recent work on quantum critical magnets.
\end{abstract}

\section{The Order Parameter Fluctuation Theory of Quantum Critical Points
and its Breakdown}

Phase transitions come in different varieties. 
Generically, they are characterized by the onset of 
an order parameter. A classical critical point, occurring at a 
finite temperature phase transition of second-order,
is described in terms 
of a coarse-grained theory of spatial, but time-independent,
fluctuations of the order parameter~\cite{Ma}. Such a description
also serves as the basis to categorize the universality classes
of critical points. Quantum critical points (QCPs) take place at
zero temperature. They differ from their classical 
counterparts in that the static (classical) and dynamic (quantum)
fluctuations are mixed and both have to be incorporated 
in the critical theory. It is, however, standard to assume 
that they too can be described in terms of fluctuations of 
the order parameter: the only distinction
being that the fluctuations are not only in 
space but also in (imaginary) time~\cite{Hertz,Sachdev-book}.

\begin{figure}
\includegraphics[width=0.8\textwidth]{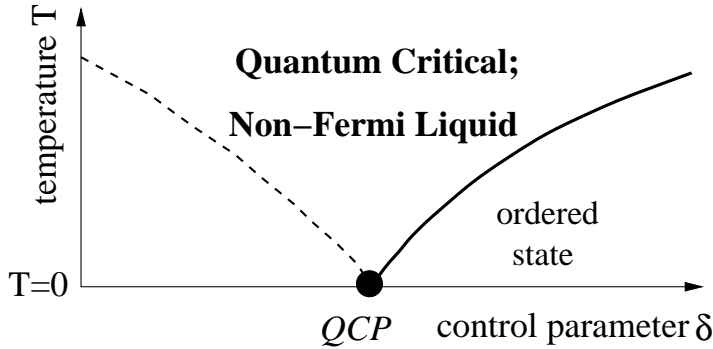}
\caption[]{Schematics of quantum phase transitions in strongly correlated
metals. Quantum criticality leads to non-Fermi liquid excitations which,
in turn, become a part of the quantum-critical spectrum and give rise to
new types of quantum-critical point.}
\label{qcp}
\end{figure}

It has been realized over the past few years that this picture
can break down in a qualitative fashion in quantum critical
metals~\cite{SietalNature2001,ColemanPepin2003}.
%%In its most drastic form -- and as we argue, also the 
%%most natural form -- such a breakdown is due to 
Non-Fermi liquid excitations emerge precisely at the QCP,
and they need to be kept as a part 
of the quantum-critical spectrum.
This is illustrated in Fig.~\ref{qcp}. On the one hand,
quantum criticality is the mechanism for the non-Fermi
liquid behavior. On the other hand,
the non-Fermi liquid excitations feed back and change
the universality class of the underlying QCP.
The experimental motivations have largely come from heavy fermion 
QCPs~\cite{Stewart,SietalNature2001,ColemanPepin2003}.
Discussions
%% about how emergent non-Fermi liquid excitations change the nature of a QCP,
of a similar spirit
can be found in an earlier pedagogical article~\cite{Si-APCTP}.

The notion that quantum fluctuations at a second-order
zero-temperature phase transition amount to adding extra
dimensions of purely classical fluctuations dates back to
the work on the Ising model in a transverse 
field~\cite{Pfeuty70,Sachdev-book}.
Within the renormalization group framework, 
Hertz~\cite{Hertz} carried out such a mapping for
a model of a metal undergoing a quantum spin-density-wave
(SDW) transition. He showed that the number of 
the extra dimensions
%% that quantum fluctuations bring about 
is equal to the dynamic exponent $z$.
The result~\cite{Hertz,Millis} is a $d_{eff}=d+z$
(where $d$ is the spatial dimensionality)
dimensional critical theory 
of paramagnons~\cite{Schrieffer,Doniach,Moriya,Lonzarich}.
It was not until recent years that quantum critical
metals became experimentally realized,
%%experiments became available, 
mostly in heavy fermion metals. In many cases,
the experiments contradict the Hertz picture
%% in a very qualitative fashion
~\cite{Stewart,SietalNature2001,ColemanPepin2003}
(see below).
%% for further discussions).

From a theoretical point of view, the possible breakdown of this
picture can be seen in a number of ways. First, the construction
of the order parameter fluctuation theory proceeds by integrating 
out fermions altogether. However, non-Fermi liquid excitations 
emerge exactly at the QCP (Fig.~\ref{qcp}); they go away
inside the phases on both sides of the QCP. These emergent excitations
should be considered as a part of the quantum-critical spectrum.
The resulting critical theory is then more than just fluctuations
of the order parameter.

Second, in the mapping of quantum
%%temporal 
fluctuations 
%%of a QCP 
into extra dimensions of 
classical fluctuations, one writes the partition function 
of a quantum mechanical system in terms of a summation over
configurations of classical variables in space and time:
\begin{eqnarray}
Z  ~\sim~\sum_{\rm config~in~({\bf x, \tau})}
~Z({\rm config}) .
\label{Z}
\end{eqnarray}
However, if the partition function for the individual 
configurations in space and time is not positive semidefinite,
Eq.~(\ref{Z}) would not necessarily correspond to a 
classical statistical mechanical theory in $d+z$
dimensions. For electronic systems, this has been known
through the ``fermion sign problem'' encountered in quantum 
Monte-Carlo methods~\cite{DagottoRMP}. The sign problem
also appears in quantum spin systems. Indeed, recent work
has demonstrated its effect in QCPs of two-dimensional
quantum magnets~\cite{SenthiletalScience2004} (see below).

\section{Quantum Critical Heavy Fermions}

Quantum criticality is of particular interest to the 
physics of strongly correlated metals. 
It provides a route towards non-Fermi 
liquid behavior~\cite{vonLohneysen,Custers}, and may also serve 
as a means of generating collective states such as unconventional
superconductors~\cite{Mathur,Flouquet}.
Heavy fermions have been playing 
an especially important role, for the simple reason that here
QCPs have been explicitly identified~\cite{Stewart}.

The contradiction to the paramagnon description
was initially found in the inelastic neutron
scattering experiments of heavy fermion metals near
an antiferromagnetic 
QCP~\cite{Schroder98,Stockert98,SchroderetalNature2000}.
The prediction of the paramagnon theory
is straightforward. Due to Landau damping, the dynamic 
exponent $z=2$ in the antiferromagnetic case.
The effective dimensionality of the fluctuations of the 
paramagnons, $d_{\rm eff}=d+2$, 
is either larger than or equal~\cite{Stockert98}
to $4$, the upper critical dimension.
The fixed point is Gaussian, and the frequency dependence
of the dynamical spin susceptibility is simply given
by the Landau damping~\cite{Hertz}, with an exponent
$1$. At the same time, its temperature dependence is 
controlled by the quartic coupling among paramagnons;
this coupling is dangerously irrelevant and produces
a temperature exponent that is larger than 1~\cite{Millis}.
An important corollary is that the dynamical spin
susceptibility necessarily violates $\omega/T$
scaling~\cite{Sachdev-book}.
(In special cases, integrating out fermions within the Hertz
framework makes the ocupling constants of the paramagnon
theory non-analytic~\cite{Belitz,Chubukov}. Such effects
may lead to fractional exponents in the dynamical spin
susceptibility at ${\bf q} \sim {\bf Q}$, but not $\omega/T$
scaling~\cite{Chubukov}.)

The experiments~\cite{Schroder98,Stockert98,SchroderetalNature2000},
on the other hand, observe $\omega/T$ scaling. Moreover,
the critical exponent for the frequency
and temperature dependences is fractional. 
Finally, this same fractional exponent is observed even 
at wavevectors far away from
the antiferromagnetic ordering wavevector ${\bf Q}$,
in a form~\cite{Schroder98}
\begin{eqnarray}
\chi({\bf q},\omega) = {{\rm const.} \over 
{f({\bf q}) + (-i\omega)^{\alpha} W(\omega/T)}} ,
\label{chi-q-omega}
\end{eqnarray}
where $f({\bf q})$ vanishes as ${\bf q}$ approaches the 
antiferromagnetic wavevector ${\bf Q}$ and stays non-zero
elsewhere. The fact that this same fractional exponent
$\alpha$ appears essentially everywhere in the Brillouin
zone suggests that its orgin is local, as noted 
in Ref.~\cite{Schroder98}.

%%At the microscopic level, h
Heavy fermions near a magnetic quantum
phase transition should be well-described in terms of a Kondo lattice
model:
\begin{eqnarray}
{\cal H} = 
\sum_{ ij,\sigma} t_{ij} c_{i\sigma}^{\dagger} c_{j\sigma}
+ \sum_i J_{_K} {\bf S}_{i} \cdot {\bf s}_{c,i} 
+\sum_{ ij} I_{ij} {\bf S}_{i} \cdot {\bf S}_{j} .
\label{kondo-lattice}
\end{eqnarray}
We are considering metallic systems, away from half-filling
({\it i.e.}, the number of conduction electrons is other than
one per $f$ electron).
We can define the tuning parameter of this model as
$\delta$, the ratio of the RKKY interaction to the 
bare Kondo scale $T_K^0$.
For negligible $\delta$, the Kondo effect 
dominates~\cite{Doniach77,Varma77}: the local 
moments are quenched and the excitations below some energy
scale, $E_{\rm loc}^*$, are spin-${1\over 2}$ and 
charge-${ e}$ Kondo resonances~\cite{Hewson}.
The result is a paramagnetic heavy fermion metal.
For large $\delta$, on the other hand,
the interactions between the local moments play the dominant role.
For dimensions higher than one and in the absence of strong 
geometrical frustration, the ground state is expected to be 
magnetically ordered. Typically, this would be an antiferromagnetic
metal (as can be expected when the system is not too far away
from half-filling). 

%%A number of important questions arise. Is the quantum phase
%%transition first order or second order? In the latter case,
What is the nature of the quantum phase transitions?
%%QCP? 
It was suggested in Ref.~\cite{SiSmithIngersent} that the behavior 
of $E_{\rm loc}^*$ can be used 
%%to characterize the nature of the quantum phase transition. 
as a characterization.
As defined earlier, this is the scale 
below which local-moments are turned into
Kondo resonances. It can also be operationally defined 
in terms of the spin damping or ``spin self-energy'' 
$M(\omega)$ (see below): $E_{\rm loc}^*$ separates the 
region ($\omega \gg E_{\rm loc}^*$) where $M(\omega)$
contains a fractional exponent and that ($\omega \ll 
E_{\rm loc}^*$) where it is linear in frequency.
Two different classes of QCP occur depending on
whether $E_{\rm loc}^*$ stays finite or vanishes at the QCP
~\cite{SiSmithIngersent}. Subsequent microscopic 
works~\cite{SietalNature2001,SietalPRB2003,ZhuGrempelSi,GrempelSi}
have indeed shown such QCPs.
Related approaches have also been developed in
%% by other authors
Refs.~\cite{ColemanetalPRB2000,ColemanPepin2003,SenthiletalPRL2003}.

\section{Microscopic Results and Their Robustness}

\subsection{Microscopic Results}

The microscopic analysis was carried out within the extended dynamical
mean field theory of Kondo lattice
systems~\cite{SietalPRB2003,SiJPCM2003} (this approach goes beyond
the dynamical mean field theory~\cite{Georges,Metzner}; for the initial
development of the method, see Refs.~\cite{SiSmith,Chitra}).
In this approach, the spin spin-energy is taken
to be momentum-independent, which is expected to be valid 
for antiferromagnetic quantum critical metals
provided the spatial anomalous dimension $\eta =0$. 

\begin{figure}
\includegraphics[width=.8\textwidth]{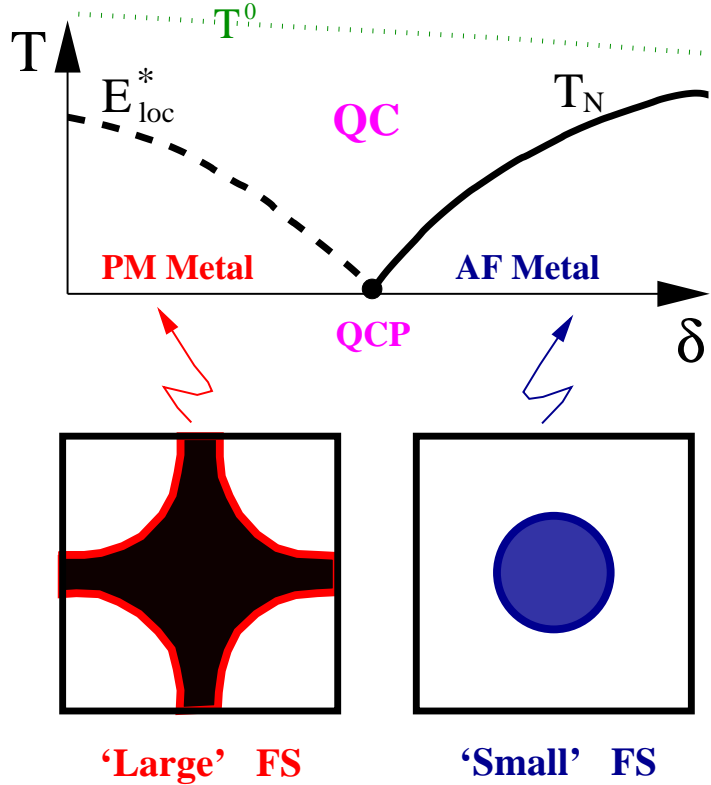}
\caption[]{Local quantum critical point. The Fermi surface undergoes
a sudden reconstruction across the QCP; within the paramagnetic
Brillouin zone, it is `large' on the heavy fermion side and `small'
on the antiferromagnetic side.}
\label{lqcp-fs-evolution}
\end{figure}

The ``local quantum critical point'' arises in this microscopic 
approach to the Kondo lattice model. The Kondo coupling is relevant
in the renormalization group sense on the paramagnetic side, 
leading to Kondo singlet formation and Kondo resonances below
a finite energy scale $E_{\rm loc}^*$. At the QCP, however,
the Kondo coupling is on the verge of becoming irrelevant. 
So $E_{\rm loc}^*$ goes to zero precisely at the QCP; 
see Fig.~\ref{lqcp-fs-evolution}.

The microscopic result also provides yet another alternative means
to define $E_{\rm loc}^*$, which appears as an infrared 
cutoff scale in the local
spin susceptibility, $\chi_{\rm loc} \equiv <\chi({\bf q},\omega)>_{\bf q}$.
On the paramagnetic metal side, $E_{\rm loc}^*$ is finite,
and $\chi_{\rm loc}$ has the Pauli behavior reflecting the 
Kondo resonances. At the QCP, on the other hand,
$E_{\rm loc}^*=0$ and $\chi_{\rm loc}$ is singular. 
The singularity is however weaker than the Curie behavior
in the non-scaling regime above the ultraviolet cutoff scale $T^0$.
($T^0$ is of the order of the single-ion Kondo temperature
$T_K^0$, and is always finite.)

The initial analytic result for the quantum-critical dynamics was 
derived with the aid of an $\epsilon-$expansion of the self-consistent
Bose-Fermi Kondo model~\cite{SietalNature2001,SietalPRB2003}.
The local spin susceptibility is
\begin{eqnarray}
\chi_{\rm loc} (\omega,T=0) = { 1 \over {2 \Lambda}}
                \ln {\Lambda \over {-i \omega}} ,
\end{eqnarray}
where the energy scale $\Lambda \approx T_K^0$. The corresponding
spin self-energy has the form
\begin{eqnarray}
M (\omega,T ) \approx -I_{\bf Q} + A ~(-i \omega)^{\alpha}W\left (
%%\omega/T)
{\omega \over T}\right ) ,
\label{spin-self-energy}
\end{eqnarray}
and the exponent is
\begin{eqnarray}
\alpha = {1 \over {2 \rho_{_I}(I_{\bf Q}) \Lambda}}
\label{fractional-exponent-alpha} .
\end{eqnarray}
Since the product $\rho_{_I}(I_{\bf Q}) \Lambda$ is expected to be of order
unity at the transition~\cite{SietalNature2001,SietalPRB2003}, 
$\alpha$ should be fractional.

The singular fluctuations seen in the local susceptibility 
reflects the embedding of the criticality of the local Kondo physics
in the criticality associated with the antiferromagnetic ordering.
The singular form might raise the concern that the local quantum 
critical point is pre-empted by magnetic ordering, which would turn
the transition first order. 
%%We have, 
It was argued that a second
order transition is natural:
for two-dimensional magnetic fluctuations, a divergent local
susceptibility necessarily accompanies a divergent peak
susceptibility (provided the spatial anomalous dimension
$\eta=0$, which is to be expected at the quantum 
transition~\cite{SietalPRB2003}).

This issue has been further addressed numerically~\cite{ZhuGrempelSi}.
We've considered the model with Ising anisotropy, the case with a
maximum tendency towards magnetic ordering. Indeed, the extrapolated
zero-temperature transition is second order. The extended dynamical
mean field analysis turns out
to be somewhat subtle. The Bose-Fermi Kondo model with Ising anisotropy
and in the absence of a static local field has a quantum phase
transition: while it is in a paramagnetic Kondo state when the coupling 
$g$ to the bosonic bath is weak, the phase at large $g$ contains
a finite Curie constant~\cite{ZhuSi}.
Since the existence of a Curie constant is hard to
see at high temperatures, it is important to reach sufficiently 
low temperatures in order to determine the nature of the zero-temperature
phase transition, as was done in Ref.~\cite{ZhuGrempelSi}. The continuous
nature was not seen in a numerical study at higher 
temperatures~\cite{SunKotliar}.

The quantum Monte-Carlo access to the local quantum critical point
also provides an opportunity to numerically determine the exponent
$\alpha$ defined in 
Eqs.~(\ref{spin-self-energy},\ref{fractional-exponent-alpha}).
In the Ising case, it is found~\cite{GrempelSi} to be approximately 
$0.72$.
Two observations about the fractional exponent are
worth making. First, the exponent turns out to be nearly universal
for this model: within the numerical accuracy, the value is the same
for four different sets of initial parameters. Second, there is 
limitation to an analytic study based on a rotor O(N)
generalization of the model [the physical case,
studied in the numerical work, corresponds to $N=1$].
While it captures 
the local quantum criticality 
(destruction of the Kondo effect
at the magnetic QCP), the rotor large-$N$ generalization,
in either the leading order~\cite{GrempelSi} in $1/N$
or the next-to-leading order~\cite{Pankovetal}, fails 
to capture the fractional exponent shown in the $N=1$ case.
This limitation is understood to be the result of a ``pinning''
of the critical amplitude of the local susceptibility in the
large-$N$ limit~\cite{GrempelSi}.

\subsection{Beyond microscopics}

These results have been argued~\cite{SietalNature2001,SietalPRB2003}
to be robust beyond the microscopic
approach, when the exponent $\alpha$ is fractional. 
The key point is that, for the long-wavelength spin fluctuations,
the dynamic exponent $z=2/\alpha$. For $\alpha<1$, it is internally
consistent to have the spatial anomalous dimension $\eta=0$.
The temporal fluctuations, on the other hand, contain a large
anomalous dimension reflecting the destruction of Kondo effect.

\subsection{In what sense is the QCP local?}

There are several ways to characterize the local nature of the 
QCP.

First, there is a localization of $f-$electrons at the QCP.
On the paramagnetic
side, the $f-$local moments are part of the low-energy electronic
excitations\footnote{Here we are concerned
with the low-energy behavior. At high-energy atomic scales,
there are no local moments and $f-$electrons are always a part 
of the electronic spectrum.}. This happens as a result of the
Kondo effect:
through the Kondo singlet formation, the local moments and conduction
electrons are entangled and the initially charge-neutral $f-$moments
are turned into charge$-e$ and spin$-{1\over 2}$ quasiparticle
excitations. In other words, the $f-$moments behave essentially as
delocalized electrons. Indeed, the Fermi surface is adiabatically
connected to that of a system in which the $f-$electrons are simply
taken as non-interacting electrons! In the case of one $f-$element
per unit cell, the Fermi volume is proportional to $1+x$, where $1$
counts the $f-$local moment and $x$ is the number of conduction 
electrons per unit cell. On the magnetically ordered side, on the 
other hand, the $f-$moments stay charge neutral and are not a part
of the electron fluid. The Fermi surface is that of the conduction
electrons alone, under the influence of a static and periodic magnetic
field produced by the antiferromagnetic order parameter. Even in
the paramagnetic Brillouin zone, the Fermi volume is proportional
to $x$. The discontinuous reconstruction of the Fermi surface takes
place precisely at the QCP; see Fig.~\ref{lqcp-fs-evolution}.
Exactly at the QCP, the effective electronic mass diverges 
over the {\it entire} Fermi surface.

Second, the local nature is also reflected in the magnetic dynamics.
The same anomalous exponent in the frequency and temperature 
dependences appears essentially {\it everywhere}
in the Brillouin zone. And the local spin susceptibility
is singular.

Both characterizations are symptomatic of the embedding of the
destruction of Kondo effect into the magnetic ordering transition.
The non-Fermi liquid excitations, which arise when the weight
of the Kondo resonance just goes to zero, are part of the 
quantum-critical spectrum. The critical theory must incorporate
these non-Fermi liquid excitations, and so it is 
no longer just the $(d+z)-$dimensional fluctuations of the order parameter
as in the paramagnon theory of the $T=0$ SDW transition.

\section{Experiments}

\subsection{Spin Dynamics}

The local quantum critical point is an interacting fixed point,
in contrast to the Gaussian fixed point of the paramagnon theory.
This is directly manifested in the spin dynamics.
The $\omega/T$ scaling and fractional exponent of the dynamical
spin susceptibility have, as mentioned earlier, already been
found in CeCu$_{\rm 6-x}$Au$_{\rm x}$. In addition,
the spin self-energy shown in Eq.~(\ref{spin-self-energy})
yields a dynamical susceptibility of the form given
in Eq.~(\ref{chi-q-omega}), with the same 
fractional exponent appearing in a broad region of the Brillouin zone.
(Closely-related features in the dynamics~\cite{Aronson95,MacLaughlin}
and thermodynamics~\cite{Maple} appear in the compound
UCu$_{\rm 5-x}$Pd$_{\rm x}$.)

The static bulk spin susceptibility is also expected to show the 
same fractional exponent\footnote{If the total
spin is conserved, the spin conservation law may modify the
behavior of the susceptibility near ${\bf q=0}$.}:
\begin{eqnarray}
\chi(T) = { 1 \over {\Theta + B T^{\alpha}}}
\label{chi-T}
\end{eqnarray}
This was already seen in CeCu$_{\rm 6-x}$Au$_{\rm x}$ early 
on~\cite{SchroderetalNature2000}.
Within a more limited temperature range, a fractional exponent
has also been observed in YbRh$_{\rm 2}$Si$_{\rm 2}$~\cite{Gegenwart}.
Eq.~(\ref{chi-T}) is also compatible with the data~\cite{Thompson,Nakatsuji}
in the normal state of CeCoIn$_{\rm 5}$, although the extent to which
this system is quantum critical remains to be established.

The NMR relaxation rate probes local spin dynamics, at a frequency
that is much smaller than the typical measurement temperature.
Assuming a featureless hyperfine-coupling,
the relaxation rate was predicted to contain a temperature-independent
component~\cite{SietalNature2001,SietalPRB2003}:
\begin{eqnarray}
{1 \over T_1} \sim A_{hf}^2 {\pi \over 8 \Lambda}
\label{1T1}
\end{eqnarray}
This has been subsequently observed 
in the Si-site NMR relaxation rate of\linebreak
YbRh$_{\rm 2}$Si$_{\rm 2}$~\cite{Ishida}.
For CeCu$_{\rm 6-x}$Au$_{\rm x}$,
recent measurement~\cite{Walstedt} of the Cu-site NMR
sees ${1/T_1} \sim T^{\alpha}$ instead.
It seems that the only way this result can be 
compatible with the inelastic neutron-scattering data is 
to invoke a strongly momentum-dependent hyperfine coupling 
constant such that the NMR relaxation rate is dominated by
contributions from generic wavevectors.
If that is indeed the case, the NMR experiment would 
provide additional evidence that the very same fractional
exponent appears at generic wavevectors.

%%Taken together, the spin dynamics in these systems are
%%definitely not compatible with the Gaussian paramagnon theory
%%but provide strong evidence for the local quantum critical
%%description.

\subsection{Thermodynamics}

A divergent specific heat coefficient (specific heat divided by
temperature) at the QCP
is expected from the localization of the $f-$electrons: the effective
mass diverges over the entire Fermi surface, as mentioned earlier.
However, a divergent specific heat coefficient can also
arise in some special cases~\cite{RoschPRL1999} of the 
SDW quantum critical point, due to contributions of the
so-called ``hot spots'' alone. 

A more definite probe turns out to come from thermodynamic ratios. 
Recall that, at classical critical points, 
thermodynamic quantities such as specific heat diverge
and provide the means to measure scaling exponents.
Since a QCP occurs at zero-temperature,
the third-law of thermodynamics dictates that the specific heat
has to go to zero. On the other hand, thermodynamic ratios 
can still diverge. A practical example is the Gr\"uneisen ratio -- the 
ratio of the thermal expansion, $\alpha \equiv {1\over V}{{\partial V}
\over {\partial T}}$, to the specific heat, $c_{p}$:
\begin{eqnarray}
\Gamma = {\alpha \over c_p} \sim 
{{\partial S / \partial p}
\over {T \partial S / \partial T}} .
\label{gruneisen-definition}
\end{eqnarray}
Under scaling, this ratio must diverge at a QCP~\cite{ZhuGarstRoschSi}.
Moreover, the Gr\"uneisen temperature exponent $x$, as defined by
$\Gamma_{\rm crit}(T) \sim 1/T^x$, measures the scaling dimension
of the most singular operator coupled to pressure.
It turns out that, within the paramagnon theory of an antiferromagnetic
SDW transition, $x$ is equal to $1$ (up to logarithmic corrections in
some cases).
At a local quantum critical point, on the other hand, 
$x$ can be fractional.

This divergence has now been observed experimentally~\cite{Kuchler}
in two heavy fermion compounds, YbRh$_{\rm 2}$Si$_{\rm 2}$ 
and CeNi$_{\rm 2}$Ge$_{\rm 2}$. For CeNi$_{\rm 2}$Ge$_{\rm 2}$,
the Gr\"uneisen temperature exponent is found to be equal to $1$.
For YbRh$_{\rm 2}$Si$_{\rm 2}$, it is fractional and is 
approximately $0.7$ -- this value is inconsistent with
the paramagnon theory but is compatible with
the scaling dimension ($0.66$ to the second order of the 
$\epsilon-$expansion)
of the operator that tunes the system to local quantum criticality.

\subsection{Electronic Measurements}

The dynamical and thermodynamic measurements provide compelling evidence
for the breakdown of the paramagnon theory in some of the heavy fermion
metals, and make a strong case 
for the local quantum critical picture. Ultimately, 
one would like to test the destruction of the Kondo effect directly,
and this can only be done through electronic measurements. Such evidence
is just emerging, from for example the recent Hall 
measurements~\cite{Paschen}.

\section{Broader context}

Heavy fermions represent a prototype strongly correlated system 
to study quantum critical physics. 
It is simpler than the doped Mott insulators in that a large
separation of energy scales exists: the local-moments are well
defined over a broad energy range, and the Kondo coupling is much
smaller than both the bare conduction electron bandwidth and the
atomic Coulomb interactions. 
It nonetheless is similar to the Mott-Hubbard systems in that
the strong Coulomb interactions lead to a microscopic Coulomb blockade.
The resulting projection of Hilbert space is essential to
the destruction of the Kondo effect
and the concomitant non-Fermi liquid excitations.
This raises the prospect that an inherent interplay between non-Fermi 
liquid and quantum critical physics takes place in other strongly
correlated electron systems as well: for high temperature
superconductors, such considerations are in line with the 
scaling behavior observed near the optimal doping.

It is also instructive to relate the quantum critical metal
physics with some recent work on quantum critical 
magnets~\cite{SenthiletalScience2004}. 
%%For magnetic systems, fractionalized excitations such as spinons
%%are the analog of the non-Fermi liquid excitations 
%%of metallic systems.
The transition considered in the Kondo lattice and that in the 
quantum magnet are different: in the former 
it is between a paramagnetic metal and an antiferromagnetic
metal and there is a well-defined order parameter only on one side;
in the latter the transition considered is between 
a ``valence-bond solid'' and an antiferromagnetic insulator,
both of which break translational symmetry and each has 
its distinct order parameter. Nonetheless, there are parallels
between the two cases:

\begin{itemize}

\item The non-Fermi liquid excitations of a Kondo lattice 
~$\longleftrightarrow$~
the fractionalized spin excitations --  ``spinons'' -- of a quantum magnet.

\item The Kondo singlets -- formed between the local moments and conduction
electrons -- of a Kondo lattice 
~$\longleftrightarrow$~
the valence bond -- or spin singlets 
formed between the local moments across bonds -- 
of a quantum magnet.

\item The Kondo singlet formation in the ground state
implies conventional Kondo resonances -- spin$-{1 \over 2}$ and 
charge$-e$ Landau quasiparticles -- in the excitation spectrum of
a Kondo lattice
~$\longleftrightarrow$~
the valence bond formation in the ground state is symptomatic of the 
confinement of spinons in a quantum magnet.

\item The destruction of the Kondo singlet is responsible
for the emergence of the non-Fermi liquid excitations at the
QCP of a Kondo lattice 
~$\longleftrightarrow$~
the destruction of the valence bond is responsible for the 
emergence of spinons that are not confined at the QCP of 
a quantum magnet.

\item The non-Fermi liquid excitations and the spinons
in a Kondo lattice and a quantum magnet, respectively,
are a part of their respective quantum-critical spectrum.

\item In both cases, the excitations of the two phases 
that the QCP separates are conventional.

\end{itemize}

These parallels arise in spite of the differences in the model 
and in the underlying physics. They suggest the exciting 
possibility that a breakdown of the fluctuating order
parameter theory
%%, and the emergent excitations being a part of the 
%%quantum-critical spectrum, occur 
occurs over a wide range of strongly correlated systems.

\section{Summary}

We have described some recent developments in the area of quantum
critical metals, with an emphasis on the interplay between
non-Fermi liquid physics and quantum critical behavior.
In addition to the traditional consideration that quantum
criticality leads to non-Fermi liquid behavior, the new 
realization is that the non-Fermi liquid excitations should
be treated as part of the quantum-critical theory thereby
leading to new classes of quantum phase transitions. 

In the specific case
of quantum critical heavy fermions, the non-Fermi liquid physics
is characterized by a destruction of the Kondo effect.
This results in the picture of local 
quantum criticality. Experimental evidence for this
picture has come from inelastic neutron scattering, NMR,
Gr\"uneisen ratio, and Hall effect.

Finally, general considerations and specific examples
suggest that the breakdown of the order parameter fluctuation
theory and the emergence of novel excitations at a quantum
critical point are properties relevant to a broad range 
of strongly correlated matters.

This article is based on a talk given at the DPG Spring Meeting
at Regensburg Germany in march, 2004. I am grateful 
to D.\ Grempel, K.\ Ingersent, S.\ Kirchner,
E.\ Pivovarov, S.\ Rabello, J.\ L.\ Smith, J.-X. Zhu,
and L.\ Zhu, as well as the Dresden group (particularly
P. Gegenwart, R. K\"uchler, J. A. Mydosh, S. Paschen, 
and F. Steglich), P. Coleman, M. Garst, and A. Rosch
for collaborations, many colleagues for discussions, and 
NSF Grant No.\ DMR-0090071 and the Robert A. Welch foundation
for support.

\end{document}